\documentclass[12pt,preprint]{aastex}

\def\gs{\mathrel{\raise0.35ex\hbox{$\scriptstyle >$}\kern-0.6em
\lower0.40ex\hbox{{$\scriptstyle \sim$}}}}
\def\ls{\mathrel{\raise0.35ex\hbox{$\scriptstyle <$}\kern-0.6em
\lower0.40ex\hbox{{$\scriptstyle \sim$}}}}

\shorttitle{HCN, HCO$^+$ as  dense molecular gas tracers}
\shortauthors{Papadopoulos}

\begin{document}

\title{HCN versus  HCO$ ^{+}$  as dense molecular  gas mass  tracer in
Luminous Infrared Galaxies}

\author{Padeli \ P.\ Papadopoulos}
\affil{Institut f\"ur Astronomie, ETH Zurich, 8093 Z\"urich, Switzerland}
\email{papadop@phys.ethz.ch}

\begin{abstract}
It has been recently argued that  the HCN J=1--0 line emission may not
be  an  unbiased tracer  of  dense  molecular  gas ($\rm  n\ga  10^4\,
cm^{-3}$)  in   Luminous  Infrared  Galaxies   (LIRGs:  $\rm  L_{FIR}>
10^{11}\,  L_{\odot}$)  and HCO$^+$  J=1--0  may  constitute a  better
tracer  instead (Graci\'a-Carpio  et  al.  2006),  casting doubt  into
earlier claims supporting the former as a good tracer of such gas (Gao
\& Solomon  2004; Wu et al.   2006).  In this paper  new sensitive HCN
J=4--3 observations  of four such galaxies are  presented, revealing a
surprisingly wide excitation range for  their dense gas phase that may
render the J=1--0  transition from either species a  poor proxy of its
mass.  Moreover the well-known sensitivity of the HCO$^+$ abundance on
the ionization degree of the molecular gas (an important issue omitted
from  the ongoing  discussion about  the  relative merits  of HCN  and
HCO$^+$  as  dense  gas  tracers)  may  severely  reduce  the  HCO$^+$
abundance in the star-forming and highly turbulent molecular gas found
in  LIRGs,  while  HCN  remains  abundant.  This  may  result  to  the
decreasing HCO$^+$/HCN J=1--0 line ratio with increasing IR luminosity
found in LIRGs, and casts doubts on the HCO$^+$ rather than the HCN as
a good  dense molecular gas tracer.   Multi-transition observations of
both  molecules are  needed  to  identify the  best  such tracer,  its
relation  to ongoing  star  formation,  and constrain  what  may be  a
considerable range of dense gas~properties in such galaxies.

\end{abstract}


\section{Introduction}

The HCN and HCO$^+$ molecules are the most abundant H$_2$ mass tracers
after  CO, whose  much higher  dipole moments  ($\mu _{10}  \sim 2.98,
3.92$\, Debye for  HCN, HCO$^+$ J=1--0 versus $\mu  _{10} =0.11$ Debye
for  CO J=1--0)  makes their  transitions excellent  tracers  of dense
molecular  gas in  galaxies.  This  is  due to  critical densities  of
rotational  transitions  being  $\rm  n_{crit}\propto \mu  ^2  \nu  ^3
_{J+1\, J}$  (for optically thin  lines at frequency $\rm  \nu _{J+1\,
J}$), allowing the HCO$^+$ and HCN lines to trace $\sim 100-500$ times
denser gas  than corresponding  (in rotational level)  CO transitions.
Early pioneering studies of the  dense molecular gas in galaxies using
HCN  and HCO$^+$  transitions  (Nguyen-Q-Rieu et  al.  1992;  Solomon,
Downes, \&  Radford 1992; Paglione  et al.  1997), have  been recently
followed  by surveys of  large galaxy  samples in  HCN J=1--0  (Gao \&
Solomon 2004a, 2004b), made possible by major advancements in receiver
sensitivity in large  mm/sub-mm telescopes.  Rotational transitions of
CS  are also  important dense  gas mass  tracers (e.g.   Plume  et al.
1997;  Shirley et al.   2003), but  are $\sim  2-6$ times  weaker than
those of  HCN (Helfer  \& Blitz 1993;  Paglione et al.   1995).  Thus,
until  the commissioning  of the  next generation  of  mm/sub-mm radio
telescope arrays, the  HCN and HCO$^+$ rotational lines  are likely to
remain  the  dense gas  mass  tracers  of  choice, especially  in  the
extragalactic domain.

The most  prominent and intriguing  result from the recent  HCN J=1--0
surveys is a nearly constant  star formation efficiency {\it per dense
molecular   gas   mass},   manifesting   itself  as   a   tight   $\rm
L_{FIR}-L_{HCN}$    correlation   and    a   nearly    constant   $\rm
L_{FIR}/L_{HCN}$  versus $\rm  L_{FIR}$ in  Luminous  Infared Galaxies
(LIRGs).   In those  systems  their IR  luminosity  ($ \rm  >10^{11}\,
L_{\odot}$) is  powered by starbursts that  are particularly prominent
in   the  Ultra   Luminous  Infrared   Galaxies  (ULIRGs)   with  $\rm
L_{IR}>10^{12}\, L_{\odot  }$ (e.g.  Genzel  et al.  1998),  while the
luminosity of  the HCN J=1--0  line ($\rm n_{crit}\sim  2\times 10^5\,
cm^{-3}$) is used  as a proxy for the dense molecular  gas mass.  In a
recent  paper Wu et  al.  (2006)  confirmed the  $\rm L_{FIR}-L_{HCN}$
linear  correlation and  extended it  down  8 orders  of magnitude  to
individual Giant  Molecular Clouds (GMCs)  found in the  Galaxy, while
locating  also its  breakdown at  $\rm L_{IR}<~10^{4.5}\,  L_{\odot }$
where the corresponding GMC masses become so small that the top of the
IMF (responsible for  the bulk of the FIR  luminosity per GMC) becomes
undersampled.  If true, such  a universal star formation efficiency of
the dense molecular  gas allows a common frame  for understanding star
formation and its  relation to the molecular gas  across cosmic epoch,
and ties this process to an obscuration-free indicator, the rotational
lines of  HCN.  The  stakes for identifying  dense molecular  gas mass
tracers in  galaxies and  their relation to  star formation  have been
recently raised by  the detection of HCN transitions  in starbursts at
high redshifts  (Solomon et  al.  2003; Wagg  et al.  2005),  but also
with recent  work casting doubt  on the reliability  of HCN as  such a
tracer and suggesting the transitions  of the molecular ion HCO$^+$ as
an alternative (Graci\'a-Carpio et al.  2006).

In  this  work new  HCN  J=4--3  ($\rm  n_{crit}\sim 8.5\times  10^6\,
cm^{-3}$) observations of four  prominent LIRGs are presented and used
to demonstrate  a surprisingly wide  range of the  physical conditions
for  the  dense molecular  gas  fueling  their  starbursts.  A  simple
corollary  of this  is that  the J=1--0  transition of  either  HCN or
HCO$^+$ may yield  very unreliable estimates of the  dense gas mass in
such  galaxies.   Moreover,   well-known  effects  particular  to  the
molecular  ion  chemistry of  HCO$^+$,  are  used  to argue  that  its
abundance  can be  greatly reduced  in the  ISM environments  found in
LIRGs.   This can  be partly  or  fully responsible  for a  decreasing
HCO$^+$/HCN J=1--0 line ratio  with IR luminosity observed recently in
LIRGs  by  Graci\'a-Carpio  et  al., and  calls  for  multi-transition
observations of  both molecules  to discern the  degree in  which they
trace the same  dense gas phase, its excitation  conditions, mass, and
relation to the often spectacular starbursts found in such galaxies.

\section{Motivation and observations}

 Early studies  of the dense molecular  gas in LIRGs  using HCN J=1--0
observations  were the first  to suggest  a potentially  constant star
formation  efficiency  per dense  gas  mass  (Solomon  et al.   1992).
Recognizing the importance of the  dense gas (defined as gas with $\rm
n(H_2)\geq 10^4\, cm^{-3}$) as the direct ``fuel'' of their prodigious
star formation an HCN and  CO, $ ^{13}$CO multi-transition line survey
was initiated with  the James Clerk Maxwel Telescope  (JCMT) in Hawaii
(US), and the IRAM 30-m telescope  at Pico Veleta (Spain) for a sample
of  30  LIRGs.   Once  completed  and  combined  with  data  from  the
literature this survey will yield a database of CO J=1--0, 2--1, 3--2,
4--3, HCN J=1--0,  3--2, 4--3, and at least  one $ ^{13}$CO transition
for all  the galaxies in the  sample.  The hereby  reported HCN J=4--3
and  CO  J=3--2 line  measurements  are  for  the (U)LIRGs:  Arp  220,
Arp~193, NGC 6240 and the  ULIRG/QSO Mrk 231, galaxies whose large HCN
J=1--0 line luminosities (larger than  the CO J=1--0 luminosity of the
Milky Way) were the first to  be measured for this class of objects by
Solomon et al.  (1992).

The observations  were conducted with the  15-m JCMT\footnote{The JCMT
is  operated by the  Joint Astronomy  Center on  behalf of  the United
Kingdom Particle  Physics and Astronomy Research  Council (PPARC), the
Netherlands  Organisation for  Scientific Research,  and  the National
Research Council of Canada.} during several periods starting from July
1999  up to  January 2006,  with  the receiver  B3 tunned  SSB to  the
frequencies  354.734\,GHz (HCN J=4--3)  and 345.796\,GHz  (CO J=3--2).
The Digital  Autocorrelation Spectrometer (DAS) was  the backend used,
set at its widest $\sim  1.8$ GHz ($\rm \sim 1520-1560\, km\, s^{-1}$)
bandwidth, except for  the HCN J=4--3 measurements of  Mrk 231 and Arp
193 where a dual-channel mode (two orthogonal polarizations) with $\rm
\sim 920\, MHz$ ($\rm \sim  777\, km\, s^{-1}$) bandwidth was used for
increased sensitivity.   Rapid beam switching with  frequencies of 1-2
Hz at  a beam  throw of $30''-60''$  (Az) produced  exceptionally flat
baselines,  and pointing  checks every  hour left  $\la 3''$  (rms) of
pointing residuals for  the $\sim 14'' $ (HPBW)  beam. Observations of
Mars and Uranus were used  to obtain the aperture efficiency, found to
be within the  expected $\sim 10\%$ of the nominal  value of $\rm \eta
_a=0.53$.   Finally frequent observations  of spectral  line standards
such as  OMC1, W75N  and W3(OH) with  high S/N  were made in  order to
ensure  the   proper  overall   line  calibration  and   estimate  its
uncertainty ($\sim 15\%$).

All  the  spectra   are  shown  in  Figure  1,   and  the  HCN  J=4--3
 velocity-integrated line flux densities were estimated from

\begin{equation}
\rm \int  _{\Delta V}  S_{\nu } dV  = \frac{8 k_B}{\eta  _a \pi
D^2}\int  _{\Delta V} T^* _A dV=  \frac{15.6 (Jy/K)}{\eta _a}
\int _{\Delta V} T^* _A dV,
\end{equation}

\noindent
(D=15\,m, point  sources assumed). These  fluxes, along with  those of
the  HCN J=1--0 line  obtained from  the literature,  can be  found in
Table 1.

\section{Dense gas in LIRGs: a surprising range of excitation}

The deduced $\rm r_{43}=(4-3)/(1-0)$ HCN ratios (Table 1) {\it imply a
surprisingly wide range of physical conditions for the dense molecular
gas phase,} with the sub-thermal  value of $\rm r_{43}=0.27$ found for
the ULIRG/QSO Mrk\,231  with the largest HCN J=1--0  luminosity of the
four, while $\rm  r_{43}\sim 1$ in Arp~220 implies  a well-excited and
dense gas phase.   In Arp~193 its weak HCN  J=4--3 line emission ($\ga
100$ weaker  than its  CO J=3--2 line)  remains undetected down  to an
impressively low limit, corresponding to $\rm r_{43}\la 0.12 $.  Large
differences in the  physical conditions of the dense  molecular gas in
galaxies with otherwise similar FIR  and low-J CO line luminosities is
well  known for systems  with $\rm  L_{FIR}\sim 10^{10}\,  L_{\odot} $
(Jackson  et  al.  1995).   The  present  work  finds such  excitation
variations to be even larger  in LIRGs with $\sim 10-100$ times larger
star  formation  rates  (and   presumably  larger  still  supplies  of
dense~gas).

 Such a large excitation range  can be detrimental when either the HCN
or HCO$^+$  J=1--0 line luminosity  is used to obtain  dense molecular
gas~mass using a standard proportionality~factor as recently advocated
by Gao \&  Solomon (2004a).  To briefly illustrate  this point a Large
Velocity Gradient  (LVG) code with HCN collisional  rates adopted from
LAMDA\footnote{Leiden      Atomic     and      Molecular     Database:
http://www.strw.leidenuniv.nl/$\sim      $moldata/\,(Sch\"oier      et
al.~2005)} is  used to  find the physical  states compatible  with the
extreme values of $\rm r_{43}\leq 0.12$ (Arp 193), and $\rm r_{43}\sim
1$ (Arp  220).  The  search is further  constrained by  $\rm T_{k}\geq
T_{dust}$ expected for the thermally decoupled gas and dust reservoirs
($\rm  T_{kin}\rightarrow  T_{dust}$ only  in  the  densest $\rm  n\ga
10^{4-5}\,cm^{-3}$,  most FUV-shielded  and  quiescent regions  inside
GMCs),   where  $\rm   T_{dust}(Arp\,  220)\sim   45\,  K$   and  $\rm
T_{dust}(Arp\,193)\sim 30\, K$; (Lisenfeld, Isaak, \& Hills~2000).

For  Arp~193  typically $\rm  n(H_2)\sim  (1-3)\times 10^4\,  cm^{-3}$
(e.g.  for $\rm T_k=30-40\, K$  and $\rm T_k=60-65\, K$), but can also
be as low  as $\rm \sim 3\times 10^3\,  cm^{-3}$ (for $\rm T_k=70-75\,
K$) and  even $\rm  \sim 10^2\, cm^{-3}$  (for $\rm T_k\geq  80\, K$).
Thus the  HCN line  emission in  Arp 193 {\it  is compatible  with the
complete absence of  a dense and massive molecular  gas phase}, and in
such a  case HCN  J=1--0 would be  tracing the  same gas phase  as the
low-J  CO transitions.   This  is  manifestly not  the  case for  $\rm
r_{43}\sim  1$, which  typically implies  $\rm  n(H_2)\sim (1-3)\times
10^5\, cm^{-3}$ (for $\rm T_{k}=40-95\, K$).  It must be noted that in
al LVG solutions  HCN J=1--0 remains optically thick,  a result easily
demonstrated using simple LTE arguments where,

\begin{equation}
\rm \frac{\tau  _{10}(HCN)}{\tau _{10}(CO)}=
\frac{\nu _{10}(HCN)}{\nu _{10}(CO)}\left[\frac{\mu _{10}(HCN)}{\mu  _{10}(CO)}\right]^2
\left(\frac{1-e^{-h\nu _{10}(HCN)/k_B T_k}}{1-e^{-h\nu _{10}(CO)/k_B T_k}}\right)
\left[\frac{HCN}{CO}\right],
\end{equation}

\noindent
($\rm h\nu  _{10}(CO)/k_B\sim 5.53\, K$,  $\rm h\nu _{10}(HCN)/k_B\sim
4.25\, K$).  For a typical $\rm [HCN/CO]\sim 2\times 10^{-4}$ and $\rm
T_{k}\sim  (15-60)\,  K$,  $\rm  \tau  _{10}(HCN)\sim  0.1\times  \tau
_{10}(CO)>1$ (since $\rm \tau _{10}(CO)\gg  1$ for the dense gas phase
traced  by  HCN).  Similar  results  are  obtained for  HCO$^+$(1--0),
reflecting the fact that their large dipole moments offset their lower
abundances relative to CO,  and keep their J=1--0 transition optically
thick for the  dense molecular gas.  Observations of  HCN, H$ ^{13}$CN
and HCO$^+$, H$ ^{13}$CO$^+$  (Nguyen-Q-Rieu et al.  1992; Paglione et
al.~1997) offer further  support for a mostly optically  thick HCN and
HCO$^+$ J=1--0 line emission.

It must  be noted that LVG  modeling of line ratios  of molecular line
emission emerging from entire galaxies  yields only a crude average of
the prevailing physical  conditions of the molecular gas,  even if one
were  to assume  the entire  emission  to be  reducible to  that of  a
typical  Galactic GMC.   Indeed,  well-known density-size  hierarchies
found in such clouds, where  $\rm \langle n(R) \rangle \propto R^{-1}$
(R is the  cloud or cloud sub-region size,  e.g.  Larson 1981), reduce
the LVG  solutions for  densities as mere  approximations of  the mean
density of  the cloud regions where  these are large  enough to excite
the transitions used.

 The mass of the HCN/HCO$^+$-emitting  dense gas can be estimated in a
manner akin to  that used for total molecular  gas mass estimates from
the $ ^{12}$CO  J=1--0 line luminosity since the  same arguments about
optically  thick   line  emission   emanating  from  an   ensemble  of
self-gravitating, non-shadowing (in  space or velocity), clouds remain
applicable (e.g.  Dickman, Snell, \& Schloerb, 1986). Thus

\begin{equation}
\rm     M_{dense}(H_2)\approx     2.1    \frac{\sqrt{\langle n(H_2)\rangle }}{T^{(10)} _{b,x}}\,
\left(\frac{L}{K\, km\, s^{-1}\, pc^2}\right)\, M_{\odot},
\end{equation}

\noindent
 where   $\rm  T^{(10)}   _{b}$  is   the  emergent   line  brightness
temperature,  and $\rm L=\int  \int T^{(10)}  _{b} da  dV$ for  HCN or
HCO$^+$ J=1--0 integrated over the entire velocity profile and area of
the source  (e.g.  Radford, Solomon,  \& Downes 1991a).  From  the LVG
solutions  the  coefficient  in Equation  3  is  found  to be  $\rm  X
_{HCN}\sim 10-30$ for  the high excitation gas in  Arp~220, but $\rm X
_{HCN}\sim  15-100$ for  Arp~193, which  may  lack a  dense {\it  and}
massive gas phase altogether (i.e.   there are LVG solutions with $\rm
n<10^4\,  cm^{-3}$).  Thus $\rm  M_{dense}(H_2)$ estimates  using only
HCN or HCO$^+$ J=1--0 line luminosities can be uncertain by factors of
$\sim 10$,  and results  based on them  must be revisited  when better
constraints on the excitation of  the dense gas become available.  For
example,  much  of the  significant  scatter  around  a constant  star
formation  efficiency  per  dense   gas  mass  (approximated  by  $\rm
L_{FIR}/L_{HCN}$) versus  star formation rate  ($\rm \propto L_{FIR}$)
found by Gao \& Solomon~(2004b), could be due to the potentially large
range of the HCN excitation revealed here. {\it The case of Arp 193, a
prominent  LIRG ($  L_{FIR}\sim 4\times  10^{11}\, L_{\odot}$)  with a
large HCN  J=1--0 line  luminosity, in which  a massive and  dense gas
phase  may not  even be  present,} demonstrates  how  singularly wrong
results can be obtained for  the gas mass at $\rm n\geq 10^4\,cm^{-3}$
in such systems if the simple approach of Equation 3 were to be used.

This stands in contrast to the $\rm X_{CO}$ factor used throughout the
literature to obtain total H$_2$ gas mass in LIRGs from the $ ^{12}$CO
J=1--0  line luminosity  (e.g. Tinney  et  al.  1990;  Sanders et  al.
1991).   The  latter line  remains  well  excited  under most  typical
conditions found in GMCs ($\rm n_{cr}\sim 400/\tau _{10}\, cm^{-3}\sim
(100-200)\, cm^{-3} $), a notion supported by CO line ratios that vary
over  a much smaller  range of  values (e.g.   Braine \&  Combes 1992;
Devereux et al.  1994; Dumke et al.  2001; Narayanan et al.  2005) and
a $\rm  X_{CO}$ found to  be robust within  factors of $\sim  2$ (e.g.
Young \& Scoville 1991).  Its  applicability is hindered only in metal
poor,  FUV-intense environments  where  CO but  not H$_2$  dissociates
(e.g.  Maloney \&  Black 1988; Israel et al.   1993, Israel 1997), and
for  highly excited, non-virialized,  gas found  in ULIRGs.   In the
latter  case  an $\rm  X_{CO}$  factor  remains  applicable, but  with
smaller values than those in the Galactic (and mostly virialized) GMCs
(Solomon et al. 1997; Downes \& Solomon 1998).

\section{HCN versus HCO$^+$ as a dense gas mass tracer in LIRGs}

Since the HCN and HCO$^+$ J=1--0  lines in the dense gas phase usually
have $\tau _{10}>1$,  estimates of its mass do  not explicitely depend
on the  specific abundances  (cf. Equation 3).   These are then  of no
great  importance, and  it could  be argued  that either  molecule can
trace the dense  gas mass equally well.  However,  in a manner similar
to the mostly FUV/metallicity-regulated  extent of the optically thick
CO J=1--0  emission relative  to the total  size of a  molecular cloud
(e.g. Pak et al. 1998; Bolatto, Jackson, \& Ingalls 1999), the HCN and
HCO$^+$ abundances partly determine the fraction of the dense gas that
can  be traced  by  their  luminous and  optically  thick J=1--0  line
emission within a  given GMC (with the level  of line excitation being
the other determining factor).

In the  dense cosmic-ray-dominated regions  of GMCs where much  of the
HCN and HCO$^+$ line  emission originates, the CR-induced formation of
$\rm  H^+  _3  $  is  the  critical initiator  of  the  networks  that
eventually yield  these two molecules.  The main  reason for expecting
$\rm [HCN/HCO^+]\ga  1$ for the dense  gas in such  regions stems from
the fact that HCN, being neutral, can remain abundant while HCO$^+$ as
a  molecular ion  is removed  via recombination  with  free electrons.
Hence,  {\it  while  both  HCN   and  HCO$^+$  are  sensitive  to  the
CR-produced  abundance of  $  H^+  _3$, HCO$^+$  is  in addition  very
sensitive to the ambient free electron abundance x(e),} and even small
increases  of the  latter  can  lead to  its  severe depletion.   This
important  point  has  been  neglected  from  the  ongoing  discussion
regarding the  relative merits of  HCN and HCO$^+$ as  effective dense
gas mass tracers (e.g. Graci\'a-Carpio et al.~2006).

This $\rm  HCO^+$ abundance sensitivity  to the ambient $\rm  x(e)$ is
well-known and  used in  studies of dense  cloud cores  throughout the
literature to obtain the latter using the former (e.g.  Wotten, Snell,
\& Glassgold  1979; Caselli et  al.  1998).  Balancing  the CR-induced
$\rm  H^+  _3$  creation   and  destruction  (via  recombination  with
electrons and HCO$^+$ formation) rates, and those for HCO$^+$, created
by $\rm  H^+ _3 +  CO\rightarrow HCO^+ +  H_2$, and removed  via e$^-$
recombination and charge transfer  reactions with metals (here assumed
negligible), yields

\begin{equation}
\rm \left[\frac{HCO^+}{CO}\right]=
\frac{k(H^+ _3, CO)\, \zeta _{CR}}{k _{HCO^+}\, k_{e}\, x(e)^2\, n(H_2)},
\end{equation}

\noindent
(e.g.  Rohlfs  \& Wilson 1996)  where $\rm \zeta _{CR}\,  (s^{-1})$ is
the CR flux, and $\rm x(e)=n_e/n(H_2)$ is the free electron abundance.
For $\rm  k(H^+ _3, CO)=1.7\times  10^{-9}\, cm^3\, s^{-1}$  (the $\rm
H^+ _3-CO$ reaction rate; Kim  et al. 1975), and $\rm k_{e}=1.26\times
10^{-6}  T^{-1/2}  _k\, cm^3\,  s^{-1}$  (McCall  et  al. 2003),  $\rm
k_{HCO^+}=6\times 10^{-6} T^{-1/2} _k\,  cm^3\, s^{-1} $ (the $\rm H^+
_3$ and $\rm HCO^+ $ dissociative recombination rates with~e$^-$),

\begin{equation}
\rm \rm \left[\frac{HCO^+}{CO}\right]\sim 2.25\times 10^{-5}\, T_k\, 
\left(\frac{\zeta _{CR}}{10^{-17}\, s^{-1}}\right)\left[\frac{x(e)}{10^{-7}}\right]^{-2}
\left[\frac{n(H_2)}{10^4\,cm^{-3}}\right]^{-1}.
\end{equation}

\noindent
In   dense   cores   in    Galactic   GMCs   $\rm   x(e)\sim   5\times
10^{-9}-1.5\times  10^{-7}$ (Langer  et al.   1985; Li  et  al.  2002)
which, for $\rm n(H_2)\sim  5\times 10^4\, cm^{-3}$, $\rm T_k\sim 15\,
K$, and  $\rm \zeta _{CR}\sim  3\times 10^{-17}\, s^{-1}$  yields $\rm
[HCO^+/CO]\sim 0.9\times 10^{-4}-0.08$,  reflecting its sensitivity on
the  recombination   with  electrons   even  in  the   rather  uniform
environment of  dense cores inside  Galactic GMCs (the  omitted charge
transfer reactions of HCO$^+$  with metals make these estimates strict
upper~limits).

However, the dark, CR-dominated, dense regions in GMCs (where the bulk
of  the  stars  corresponding  to  a  normal IMF  forms)  may  not  be
representative  of  the  typical  ISM environments  in  LIRGs.   There
intense FUV  and even X-ray  (when AGNs are present)  radiation fields
may   dominate   and   shift   the   dense   gas   chemistry   towards
photon-dominated  rather  than  CR-dominated processes.   Nevertheless
models  of  such photon-dominated  regions  (PDRs)  yield mostly  $\rm
[HCN/HCO^+]> 1$, and observations of PDRs such as IC 63, NGC 2023, and
the Orion Bar seem to confirm this by finding $\rm N(HCN)/N(HCO^+)\sim
1-5$ (Jansen 1995).  Here must be noted that unexpectedly weak CN line
emission  with small CN/HCN  ratios that  decrease with  increasing IR
luminosity in  LIRGs, casts doubts on  the prevelance of  PDRs for the
bulk of their molecular gas  reservoirs (Aalto 2004).  This is because
PDR models invariably predict very large CN/HCN ratios on the surfaces
of FUV-illuminated clouds (e.g.  Boger \& Sternberg 2005), making this
ratio one of the most  effective diagnostics of their presence.  Given
the large extinctions found in LIRGs ($\rm A_v\sim 50-1000$; Genzel et
al.  1998) it is possible that the FUV radiation from the newly formed
O,  B star  clusters is  effectively  absorbed almost  in situ  around
ultra-compact HII  regions, and thus  a lower ambient  radiation field
irradiates the bulk of the molecular gas in these~galaxies.

  The mostly low CN/HCN line  intensity ratios found in LIRGs, many of
which  also host  AGNs, signifies  also a  negligible  contribution of
X-ray  Dissociation Regions  (XDRs) to  the  {\it bulk}  of the  large
molecular gas reservoirs  found in these galaxies, since  in XDRs $\rm
[CN/HCN]\sim 5-10$ (Lepp \&  Dalgarno 1996).  The well-studied case of
the molecular gas in the starburst/AGN Seyfert 2 galaxy NGC 1068 shows
the  influence of  XDRs limited  to the  small fraction  of  the total
molecular gas residing close to the AGN (Tacconi et al. 1994; Usero et
al.  2004), and  in those regions one actually  finds enhanced HCN and
diminished HCO$^+$  J=1--0 line intensities (Kohno et  al.  2001). For
more  powerful AGNs  XDR  chemistry could  become  relevant for  large
fractions  of molecular  gas mass  in the  galaxies hosting  them, and
detailed work  studying its effects  on molecular abundances  can help
identify observationally  accessible signatures (e.g.   Maloney et al.
1996; Meijerink \& Spaans 2005).

\subsection{Turbulence, dense and ion-rich ISM in LIRGs:  surpressors
of the HCO$^+$ abundance}

The presence of turbulence in molecular clouds is now well established
(e.g.  Falgarone 1997), and its  strong effects to their chemistry via
the   turbulent  diffusion   of  the   ion-rich  (mostly   C$^+$)  and
electron-rich outer layers inwards have been demonstrated (Xie, Allen,
\& Langer 1995).   {\it For turbulence levels easily  attained in GMCs
in LIRGs  $\rm x(e)$ rises by  an order of magnitude,}  and HCO$^+$ in
dense cloud interiors then becomes  surpressed by almost two orders of
magnitude (Figs  2, 3  in Xie et  al.), in approximate  agreement with
Equation~5    (for     a    constant    CO     abundance).     Thicker
(C$^+$,e$^{-}$)/C-dominated zones  on GMC surfaces in  LIRGs (a result
of  potentially  larger FUV  radiation  fields)  from which  turbulent
diffusion  can draw  e$^{-}$-rich molecular  gas inwards  will further
surpress  HCO$^+$.   On the  other  hand  the  HCN abundance  will  be
enhanced by  the now  larger quantities  of $\rm C^+$  and C  in cloud
interiors  since they  facilitate efficient  HCN production  (Boger \&
Sternberg~2005).  Finally the shocks  that are expected to be frequent
in the highly  supersonic turbulent molecular gas found  in LIRGs, can
also significantly reduce the  HCO$^+$ while leaving the HCN abundance
unperturbed (Iglesias \& Silk 1978; Elitzur~1983).

 In  all cases  where  a  high $\rm  [HCN/HCO^+]$  abundance ratio  is
 expected, a  potentially larger  portion of the  dense gas  phase may
 become more luminous  through the HCN rather than  the HCO$^+$ J=1--0
 line emission and,  depending of the level of  line excitation, yield
 larger values of $\rm M_{dense}(H_2)$ via Equation 3.

\subsection{Effects favoring HCO$^+$ as dense gas tracer}

In completely FUV-shielded  environments photoionization is negligible
 and cosmic  rays will be the  sole cause of ISM  ionization, and thus
 $\rm x(e)$ and $\rm \zeta  _{CR}$ are not expected to be independent.
 Following a treatment by McKee (1989)

\begin{equation}
\rm x(e) = 2\times 10^{-7} \left(\frac{n_{ch}}{2n(H_2)}\right)^{1/2} 
\left[\left(1+\frac{n_{ch}}{8n(H_2)}\right)^{1/2} + \left(\frac{n_{ch}}{8n(H_2)}\right)^{1/2}\right],
\end{equation}

\noindent
where $\rm  n_{ch}\sim 500\, \left(r ^2  _{gd}\, \zeta _{-17}\right)\,
cm^{-3}  $ is  a characteristic  density encapsulating  the  effect of
cosmic rays  and ambient metallicity  on the ionization  balance ($\rm
r_{gd}$:  the  normalized gas/dust  ratio,  $\rm  r_{gd}=1$ for  Solar
metallicities). From Equations 5 and 6,

\begin{equation}
\rm \left[\frac{HCO^+}{CO}\right] \sim 2.25\times 10^{-4}\, T_k\, r^{-2} _{gd}\,
\left[\left(1+\frac{n_{ch}}{8n(H_2)}\right)^{1/2} + 
\left(\frac{n_{ch}}{8n(H_2)}\right)^{1/2}\right]^{-2}.
\end{equation}

In the last  expression $\rm [HCO^+/CO] $ appears  much more robust in
changes of  the ambient ISM properties  than in Equation  5 where $\rm
x(e)$  and $\rm  \zeta _{CR}$  were considered  independent.   For gas
dense  enough  to excite  the  HCO$^+$  J=1--0  line ($\rm  n_{cr}\sim
3.4\times 10^4\,  cm^{-3}$), and  quiescent conditions typical  of the
Galaxy  ($\rm  \zeta  _{-17}\sim  1-3$, $\rm  n_{ch}\sim  (500-1500)\,
cm^{-3}$), it  is $\rm n_{ch}/(8n(H_2))\ll 1$ and  $\rm [HCO^+/CO]$ is
independent  of   gas  density  and   CR  flux.   Only   in  starburst
environments  where $\zeta _{-17}=100-500$  (and thus  $\rm n_{ch}\sim
(0.5  -2.5)\times  10^5\,   cm^{-3}$),  serious  surpression  of  $\rm
[HCO^+/CO]$ can occur in gas dense enough to excite the HCO$^+$ J=1--0
transition  (e.g. $\rm  [HCO^+/CO]\sim (2-8)\times  10^{-4}$  for $\rm
n(H_2)=10^4\,  cm^{-3}$  and  $\rm  T_k=15\, K$).   However  extensive
observations of dense  cores in Galactic GMCs do  not support Equation
6, finding $\rm x(e)$ in most  such regions to be much higher (Caselli
et al.   1998). Inward turbulent  transport of the outer  cloud layers
where $\rm x(e) $ is much higher because of photoinization could

It must  be mentioned that  not all the  effects of turbulence  on the
HCO$^+$ abundance  are negative since its  intermittent dissipation in
diffuse  ($\rm   n(H_2)\sim  10^2-10^3\,  cm^{-3}$)   and  warm  ($\rm
T_{k}\sim 100-200\,  K$) molecular  gas can cause  significant HCO$^+$
abundance enhancements but this process involves only small amounts of
gas (Falgarone 2006).

Abundance ratios of $\rm  [HCN/HCO^+]>1 $ do not necessarily translate
to similar  line intensity ratios  for the corresponding (in  J level)
HCO$^+$  and HCN  transitions since  their  excitation characteristics
differ.   Their $\rm E_u/k_B$  values are  similar but  their critical
densities  $\rm  n_{crit}(HCN)/n_{crit}(HCO^+)\sim  5-7$ (for  J=1--0,
3--2 and 4--3) and this will modify and can even reverse any abundance
advantage that HCN  may have over HCO$^+$ and  make the transitions of
the latter brighter.  Indeed,  given the density gradients expected in
GMCs, even small differences in  $\rm n_{crit}$ can translate to large
differences  in the  extent of  the cloud  rendered ``visible''  via a
particular  transition.  For the  density-size relation:  $\rm \langle
n(R)\rangle \propto R^{-1}$, and  assuming gas remains ``visible'' via
a particular  molecular line out to  a radius $\rm  R_{cr}$ where $\rm
\langle n(R_{crit})\rangle \sim n_{crit}$, the cloud mass ratio probed
by HCO$^+$ and HCN (assuming equal abundances) will be

\begin{equation}
\rm \frac{M_{HCO^+}(H_2)}{M_{HCN}(H_2)}\sim \left[\frac{R_{crit}(HCO^+)}{R_{crit}(HCN)}\right]^2\sim
\left[\frac{n_{crit}(HCN)}{n_{crit}(HCO^{+})}\right]^2\sim 25-49.
\end{equation}

\noindent
These  values  are  only  indicative,  since  optical  depth  effects,
temperature,  and abundance gradients  will modify  the aforementioned
simple picture.  Only observations  of high-J lines for both molecules
can discern  the most comprehensive  tracer of dense gas  by comparing
their respective (4--3)/(1--0),  (3--2)/(1--0) line ratios as recently
done by Greve et al. 2006.

Finally irrespective  of which of the  two molecules turns  out as the
most  encompassing  tracer  of  molecular  gas at  $\rm  n\geq  10^4\,
cm^{-3}$, it  may still  include large amounts  of gas  not intimately
involved  in  the star  formation  process.   Indeed if  observational
studies  of high-mass star-forming  cores (Shirley  et al.   2003) and
recent theoretical  advances (Krumholz \&  McKee 2005) are  any guide,
molecular gas  with $\rm  n\ga 10^5\,cm ^{-3}$  is expected to  be the
true star formation fuel in  the turbulent GMCs.  Thus observations of
high-J  transitions of  large dipole  moment molecules  such  as those
presented  here, aside  from  yielding constraints  on the  excitation
conditions of the dense gas phase,  are a step closer to the true fuel
of  star formation in  LIRGs, and  a steping  stone for  revealing any
universal aspects of this process~in~galaxies.

\section{Conclusions}

In  this   work  sensitive  new   HCN  J=4--3  observations   of  four
prototypical  Luminous  Infrared Galaxies  (LIRGs)  are presented  and
combined with existing HCN J=1--0 measurements to probe the excitation
properties  of the  dense gas  ($\rm n\geq  10^4\, cm^{-3}$)  in these
remarkable objects.  The results,  along with well-known  effects that
can  severely affect  the HCO$^+$  abundance, are  used to  insert the
following  important  points  in  the  ongoing  debate  regarding  the
relative merits of HCN and HCO$^+$  lines as dense gas mass tracers in
such galaxies,

\noindent
1. The  large range  of  excitation conditions  revealed  from a  {\it
  global}  HCN  (4--3)/(1--0) ratio  varying  by  almost  an order  of
magnitude ($\sim  0.1-1$) among the  LIRGs observed here  can severely
hamper the methods advocated recently  for dense gas mass estimates in
such systems, by  rendering the HCN or HCO$^+$  J=1--0 line luminosity
and a ``standard'' conversion factor a poor proxy for that mass.

\noindent
2.  HCO$^+$, unlike HCN, is a  molecular ion and thus easily destroyed
by  recombination with free  electrons. Under  most conditions  in the
dense  gas  regions of  molecular  cloud  interiors  this yields  $\rm
[HCO^+/HCN]<  1$, especially  in environments  with  enhanced electron
abundances.   In the  turbulent  and FUV-irradiated  molecular gas  in
LIRGs such  environments are  expected to be  common, and  this effect
could  cause the low  $\rm HCO^+/HCN$  line intensity  ratios observed
recently in such galaxies, especially towards high IR luminosities.

\noindent
3. The lower critical densities  of the HCO$^+$ rotational transitions
than  the corresponding  ones (in  J-level) of  HCN will  moderate and
could even reverse the abundance advantage of the latter when it comes
to line brightness.  Thus, aside from yielding valuable constraints on
what may be  a considerable range of dense  gas excitation properties,
J=4--3, 3--2 observations for both molecules will help decide on their
relative  merits  as dense  gas  mass  tracers.   Finally such  high-J
transitions  can   potentially  trace  the  much   denser  $\rm  (n\ga
10^5-10^6\, cm^{-3}$) molecular gas,  thought to be the immediate fuel
of star formation in~LIRGs.

\acknowledgments

The   author   is   grateful   for  extensive   comments   by   Javier
Graci\'a-Carpio and  Santiago Burillo  that helped to  greatly improve
the original manuscript, and especially for pointing out their new HCN
J=1--0 measurement of Arp 193 and its large discrepancy with the value
previously reported  in the literature. Comments and  questions by the
referee were very helpful in clarifying key aspects of this work.

\clearpage

\begin{figure}[h]
%\centering

\vspace*{8cm}
\leavevmode
\includegraphics{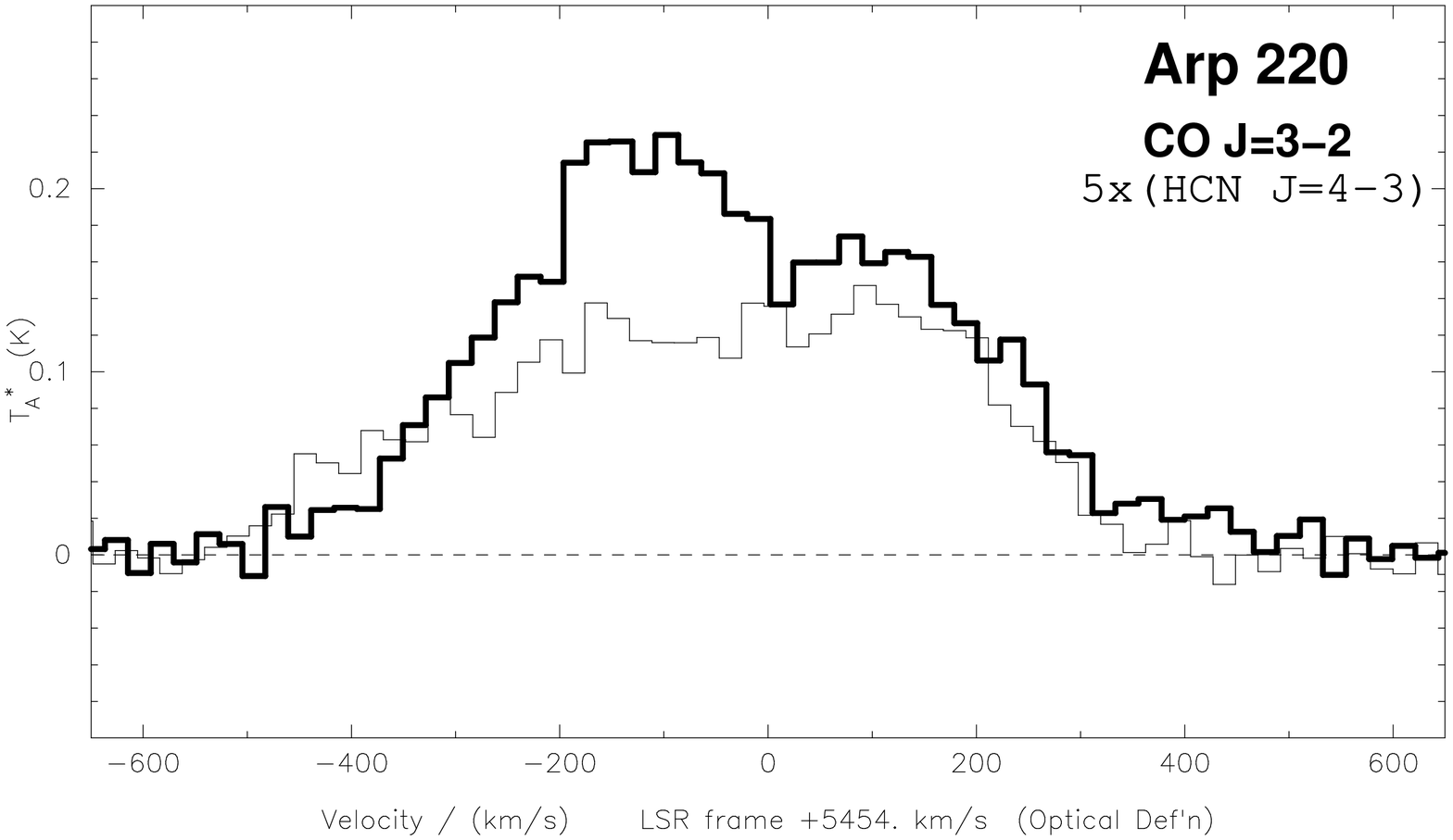}
\includegraphics{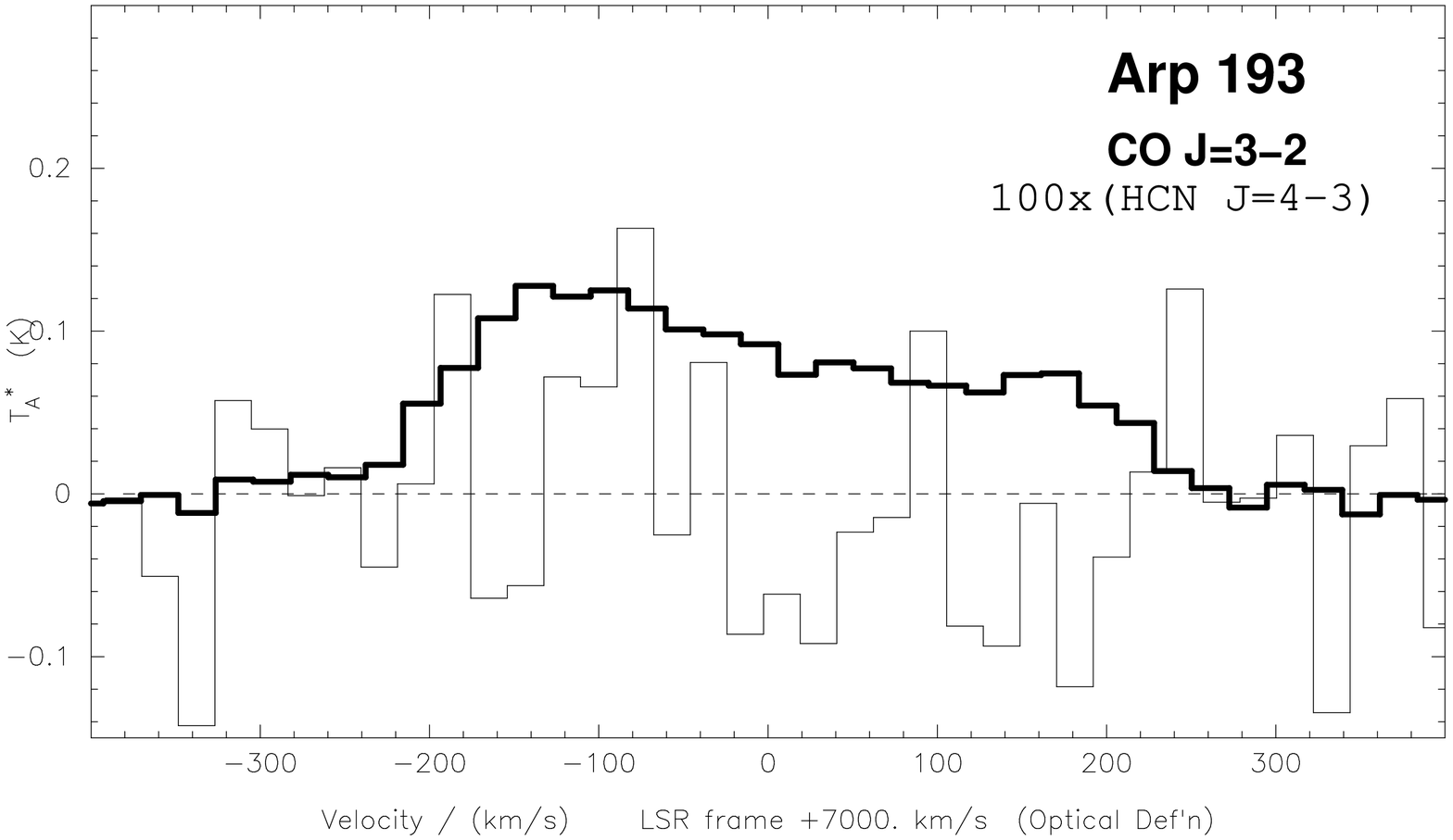}
\includegraphics{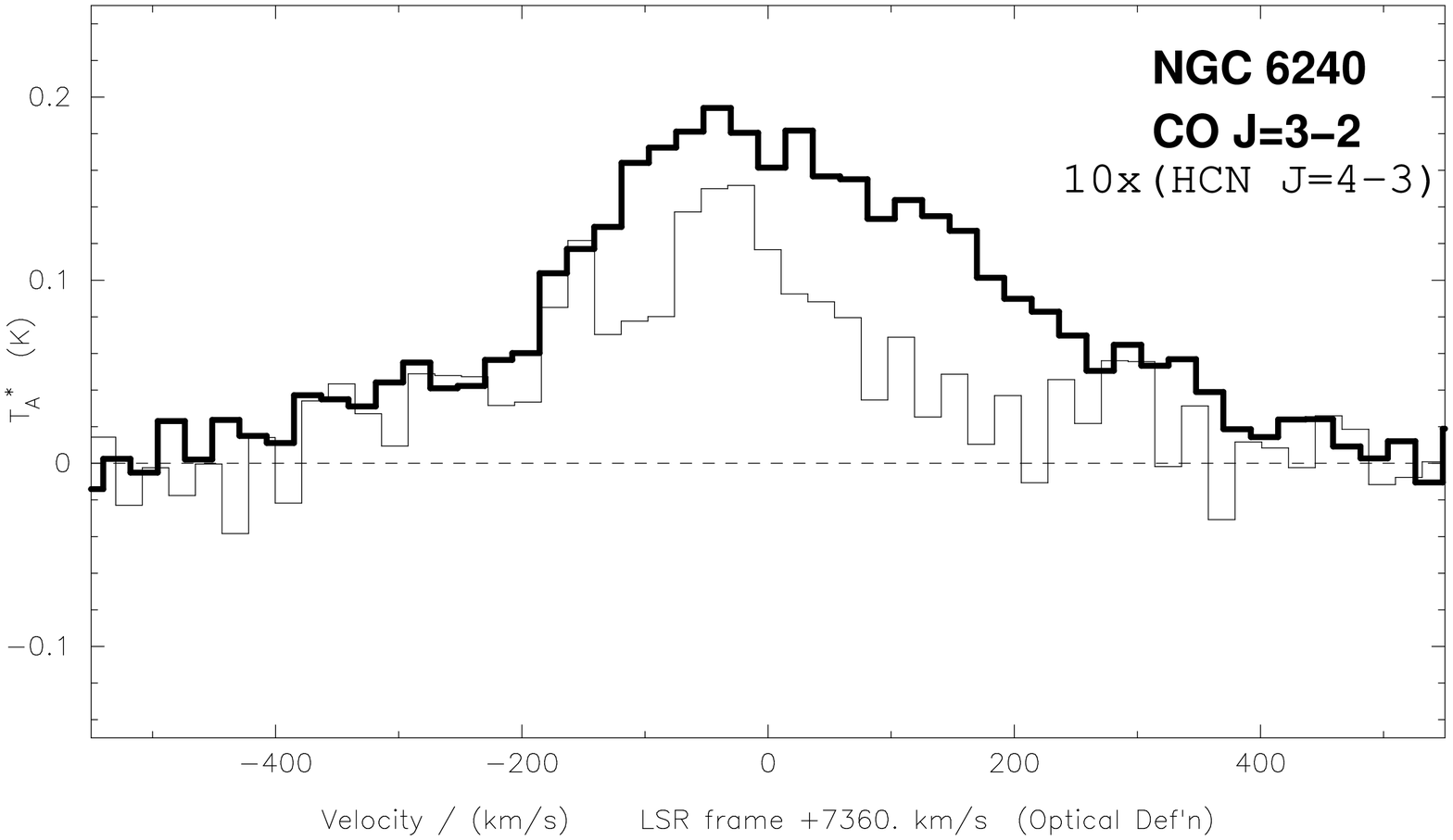}
\includegraphics{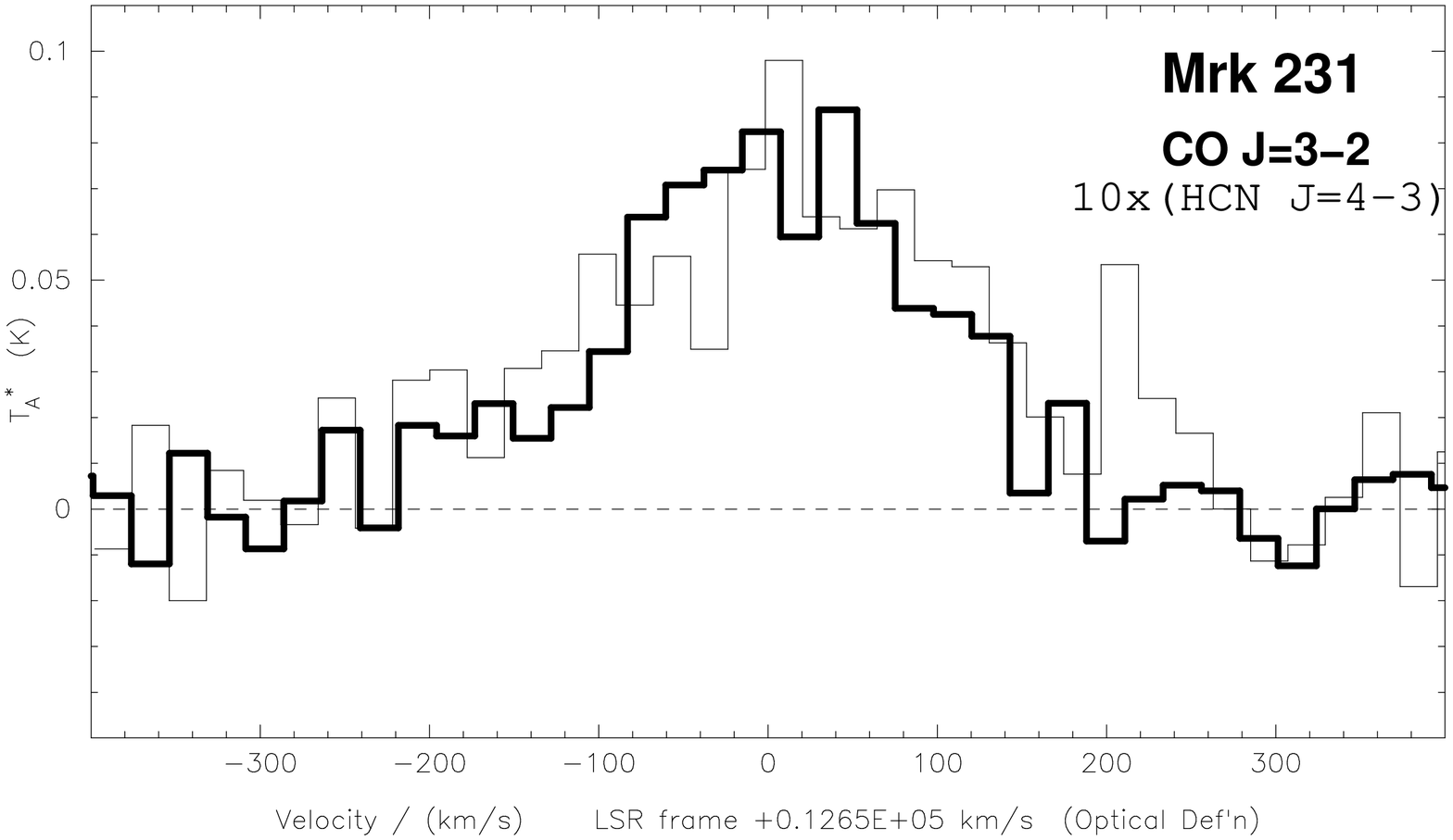}
\vspace*{5cm}
\caption{HCN J=4--3 (scaled by numbers  given in the upper right), and
CO  J=3--2 spectra  (bold line)  of the  four LIRGs  at a  common $\rm
\Delta  \nu _{ch}=25\,  MHz\,  (\sim  21\, km\,  s^{-1}$  at 350  GHz)
resolution.}
\end{figure}

\newpage

\begin{deluxetable}{lcccc}
\tablecolumns{5}
\tablewidth{0pc}
\tablecaption{HCN line intensities and ratios}
\tablehead{
\colhead{Galaxy}&\colhead{$\rm L_{IR}$\tablenotemark{a}}&\colhead{$\rm \int S_{HCN(4-3)} dV$} & 
 \colhead{$\rm \int S_{HCN(1-0)} dV$}&\colhead{$\rm r_{43}(HCN)$\tablenotemark{b}}\\
    &  $\rm (\times 10^{11}\, L_{\odot})$ &  Jy\,km\,s$^{-1}$ & Jy\,km\,s$^{-1}$ & }  
\startdata
Arp 220  & 14   & $577\pm 105$ & $\rm 36\pm 7$\tablenotemark{c} & $1.00\pm 0.25$ \\
Arp 193  & 3.7  & $\leq 10 (3\sigma) $ & $5\pm 1$\tablenotemark{d}  & $\leq 0.12$ \\
Mrk 231  & 30.3 & $65\pm 13$  & $15\pm 3$\tablenotemark{e}   & $0.27\pm 0.08 $\\
NGC 6240 & 6.1  & $130\pm 25$  & $14\pm 2$\tablenotemark{f}   & $0.60\pm 0.15 $
\enddata
\tablenotetext{a}{From Gao \& Solomon 2004b.}
\tablenotetext{b}{Brightness temperature ratios ($\rm T_b$ averaged over area/velocity).}
\tablenotetext{c}{Average from Radford et al. 1991b, and Solomon et al. 1992.}
\tablenotetext{d}{Garcia-Carpio private communication.}
\tablenotetext{e}{From Solomon et al. 1992.}
\tablenotetext{f}{From Greve et al. (2006)}
\end{deluxetable}

\end{document}